\documentclass[aps,prl,twocolumn,superscriptaddress, showpacs]{revtex4}

\usepackage{graphicx}
\usepackage{dcolumn}

\begin{document}


\title{ Joint Temporal Density Measurements for Two-Photon State 
Characterization}


\author{Onur Kuzucu}
\affiliation{Research Laboratory of Electronics, Massachusetts Institute of 
Technology, Cambridge, Massachusetts 02139, USA}

\author{Franco N. C. Wong}
\affiliation{Research Laboratory of Electronics, Massachusetts Institute of 
Technology, Cambridge, Massachusetts 02139, USA}

\author{Sunao Kurimura}
\affiliation{National Institute for Materials Science, 1-1 Namiki, Tsukuba-shi, 
Ibaraki 305-0044, Japan}

\author{Sergey Tovstonog}
\affiliation{National Institute for Materials Science, 1-1 Namiki, Tsukuba-shi, 
Ibaraki 305-0044, Japan}


\date{\today}

\begin{abstract}
We demonstrate a new technique for characterizing two-photon quantum states 
based on joint temporal correlation measurements using time-resolved 
single-photon detection by femtosecond upconversion.  We measure for the first 
time the joint temporal density of a two-photon entangled state, showing clearly  
the time anti-correlation of the coincident-frequency entangled photon pair 
generated by ultrafast spontaneous parametric down-conversion under extended 
phase-matching conditions.   The new technique enables us to manipulate the 
frequency entanglement by varying the down-conversion pump bandwidth to produce 
a nearly unentangled two-photon state that is expected to yield a heralded 
single-photon state with a purity of 0.88. The time-domain correlation technique 
complements existing frequency-domain measurement methods for a more complete 
characterization of photonic entanglement in quantum information processing. 
\end{abstract} 


\maketitle

Spontaneous parametric down-conversion (SPDC) is a powerful method for 
generating two-photon states for quantum information processing (QIP).  The 
joint quantum state can be engineered for specific QIP applications by tailoring 
its polarization, momentum, and spectral degrees of freedom.  Ultrafast-pumped 
SPDC is of great interest because a well defined time of emission is desirable 
in clocked applications such as linear optics quantum computing (LOQC) 
\cite{klm01}.  In ultrafast SPDC, spectral engineering of the two-photon state 
can be accomplished by manipulating the crystal phase-matching function and the 
pump spectral amplitude \cite{grice, giovannetti_epm} to yield unique forms of 
two-photon frequency entanglement.  For example, coincident-frequency 
entanglement with strong positive correlation between signal and idler emission 
frequencies can be used to improve time-of-flight measurements beyond the 
standard quantum limit \cite{giovannetti_nature, kuzucu}.  On the other hand, 
one can utilize a two-photon state with negligible spectral correlations to 
implement a heralded source of pure-state single photons, which can be a 
valuable resource for LOQC \cite{uren, mosley}.  

Characterizing the spectral correlations of a two-photon state can be done by 
measuring the joint spectral density (JSD) profile with tunable narrowband 
filtering of the signal and idler \cite{uren, mosley, hendrych}.  Hong-Ou-Mandel 
quantum interference \cite{HOM} is also useful for quantifying the two-photon 
coherence bandwidth and the indistinguishability of the photon pair.  However, 
the two measurements do not give the whole picture of the two-photon state. Both 
measurements are insensitive to the spectral phase and therefore cannot capture 
the time-domain dynamics unless the joint state is known to be transform 
limited. Moreover, JSD measurements in wavelength regions with low detector 
efficiency or high detector noise can be challenging due to long acquisition 
times and low signal-to-noise ratios. Frequency-domain techniques for estimating 
the spectral phase exist, but they are not simple to implement in practice 
\cite{wasilewski}.

In ultrafast optics ultrashort pulses are routinely analyzed spectrally and  
temporally, but time-domain characterization tools are not easy to implement for 
single photons. Recently we have introduced a time-resolved single-photon 
measurement technique by use of femtosecond upconversion \cite{ol_submission}.  
In this Letter we utilize this single-photon time-domain characterization method 
to measure for the first time the joint temporal density (JTD) profile of a 
two-photon quantum state. In particular, we measured directly the time 
correlations of signal-idler arrival times of ultrafast pumped SPDC under 
extended phase matching conditions \cite{kuzucu}, showing clearly that the 
coincident-frequency entangled photons were time anti-correlated.  Furthermore, 
by varying the SPDC pump spectrum, we were able to manipulate the temporal 
correlations of the signal and idler, and obtain a nearly unentangled 
(temporally) two-photon state.  This new technique can be used in conjunction 
with frequency-domain methods to provide a more complete characterization of 
single and entangled photons. 

\begin{figure}[h]
\includegraphics[width=3.2in]{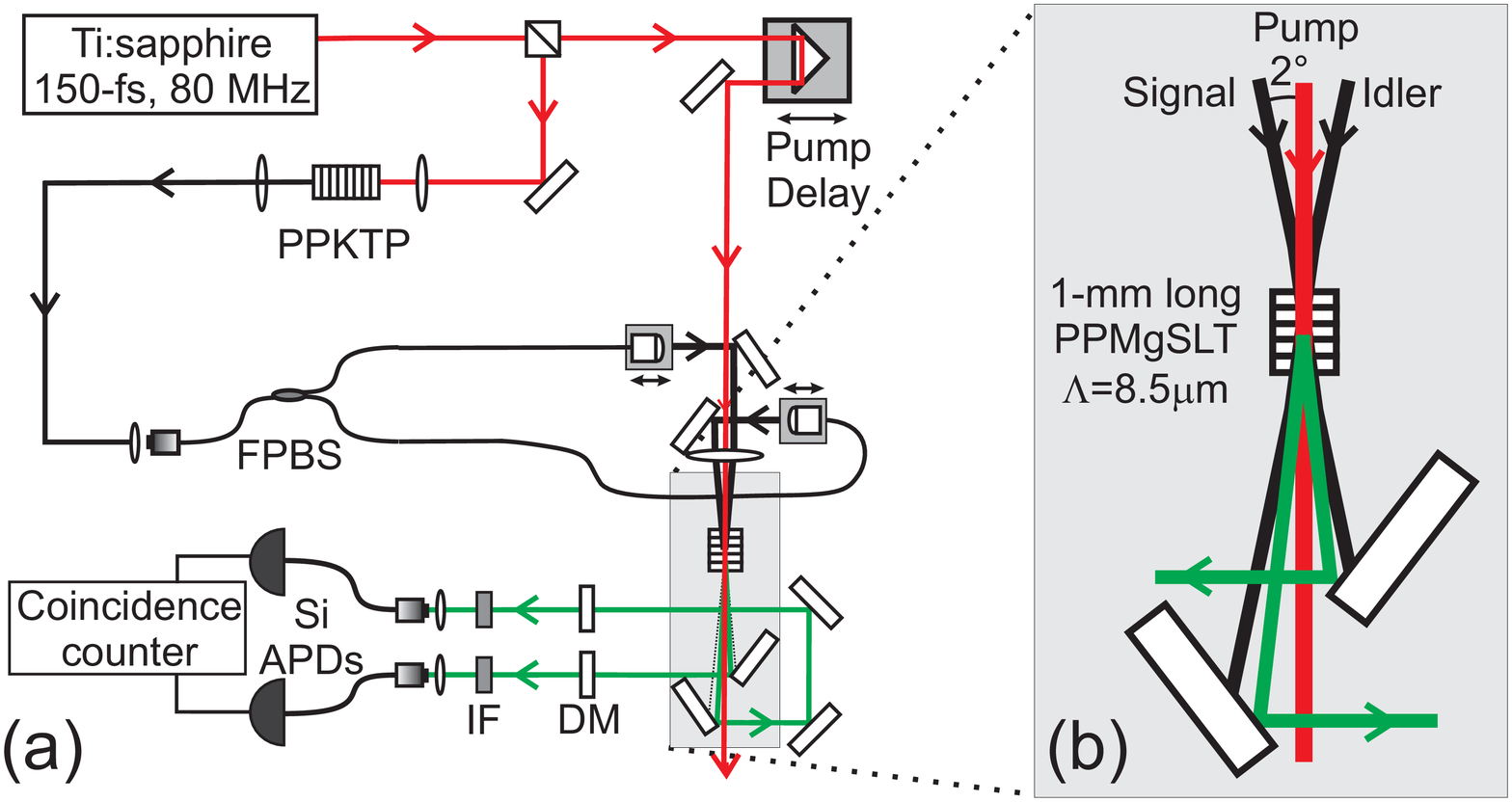} 
\caption{(Color online) (a) Synchronized upconversion and downconversion 
experiment driven by the same ultrafast pump. (b) Noncollinear phase-matching 
geometry for single-photon upconversion. IF: interference filter; DM: dichroic 
mirror; FPBS: fiber polarizing beam splitter.} \label{upconv_sketch}
\end{figure}

To properly define JTD, we first express the two-photon state in time-domain 
variables $|\Psi\rangle = \int \hspace{-0.15cm} \int d\tau_S
d\tau_I\, {{\cal A}}(\tau_S, \tau_I) |\tau_S\rangle |\tau_I\rangle$, where the 
single-photon Fock state is defined as
$|\tau_j\rangle\equiv\hat{a}^\dagger(\tau_j)|0\rangle$, for $j=S,I$. The 
temporal correlations of the signal and idler are determined by the
joint temporal amplitude, ${{\mathcal{A}}}(\tau_S, \tau_I)$, and we define the 
associated probability density, ($\left|{{\mathcal{A}}}
(\tau_S, \tau_I) \right|^2$), as the joint temporal density. Analogous to the 
frequency-domain methods, the JTD can be measured by using narrowband 
\emph{temporal} filtering and coincidence detection. For typical ultrafast SPDC 
experiments, timing resolution of $\sim$100\,fs is needed for measuring arrival 
times of single photons.  Current single-photon detectors with tens of 
picoseconds timing resolution are not suitable for this purpose. For the 
two-photon JTD measurement, we applied our recently developed time-resolved 
single-photon upconversion technique with a temporal resolution of $\sim$150\,fs 
\cite{ol_submission}. An ultrafast upconverting pump pulse was used to 
time-stamp the signal and idler arrival times, and we mapped their relative 
arrival times by varying the input delay lines independently and recording the 
coincidences between the two upconversion channels.  The coincidence statistics 
yielded the temporal structure of the two-photon state.

Our experimental setup for ultrafast type-II phase-matched SPDC and subsequent 
JTD measurement with time-resolved upconversion is shown in 
Fig.~\ref{upconv_sketch}(a). Both SPDC and upconversion were pumped 
synchronously with the same ultrafast source at 790\,nm with a 6-nm bandwidth 
and 80\,MHz repetition rate, thereby eliminating the pump timing jitter for the 
JTD measurement. We operated the PPKTP SPDC crystal under extended 
phase-matching conditions to generate a coincident-frequency entangled 
two-photon state \cite{giovannetti_epm, kuzucu}. By Fourier duality, this 
positive frequency correlation corresponded to anti-correlation in the time 
domain where the signal and idler photons with $\sim$350-fs single-photon 
coherence times were symmetrically located about the center of a $\sim$1.4-ps 
two-photon coherence time window, as measured by HOM interference \cite{kuzucu}. 
The signal and idler photons were coupled into a polarization-maintaining 
single-mode fiber and separated at a fiber polarizing beam splitter. The signal 
and idler delay lines were individually adjusted so that they arrive at the 
upconversion crystal in the same time slot as the pump pulse.  Fine tuning of 
the relative timing can be achieved with translation stages. 

We used the same setup as in Ref.~\cite{ol_submission} for time-resolved 
single-photon upconversion, briefly described here.  As sketched in 
Fig.~\ref{upconv_sketch}(b), a 1-mm long periodically poled MgO-doped 
stoichiometric lithium tantalate (PPMgSLT) crystal with a 8.5\,$\mu$m grating 
period was used for noncollinear type-0 phase-matched sum-frequency generation  
(1580\,nm + 790\,nm $\rightarrow$ 526.7\,nm).  We used the noncollinear geometry 
to implement two independent upconverters with a single crystal.  The 
single-photon beams were aligned parallel to the pump beam with $\sim$3\,mm 
lateral and $\sim$1.5\,mm vertical separation from the pump axis, and they were 
focused into the PPMgSLT crystal. The non-planar focusing configuration allowed 
us to avoid the simultaneous detection of the non-phase-matched parametric 
photon pairs that were both generated and upconverted by the pump at the
PPMgSLT crystal. Therefore, even with a finite background for singles, the 
coincidence profile shows negligible accidentals \cite{ol_submission}.
The upconverted outputs were filtered by dichroic mirrors and 10-nm passband 
interference filters, coupled into single-mode fibers and detected with 
fiber-coupled Si APDs. We recorded the singles counts and also the coincidence 
counts between the two Si APDs within a 1.8\,ns coincidence window.

\begin{figure}[h]
\includegraphics[width=3.2in]{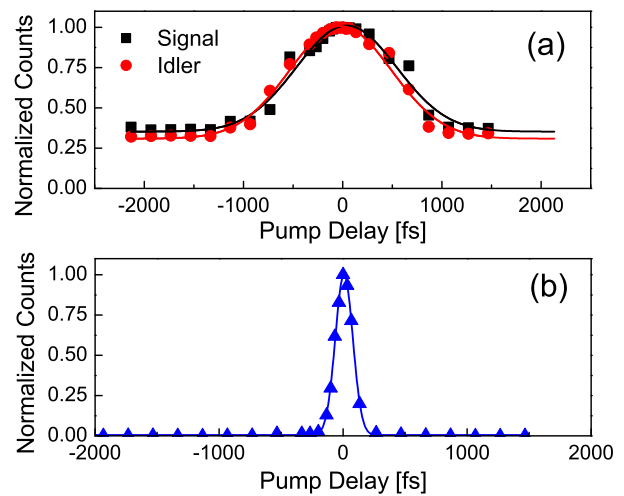} 
\caption{(Color online) Normalized singles (a) and coincidence (b) histograms by 
time-resolved upconversion. The pump pulse was scanned through collocated signal 
and idler arrival windows. Solid lines are Gaussian fits to the data.
} \label{pump_scan_a}
\end{figure}

We measured the singles and coincidences by scanning the upconversion
pump pulse delay relative to the signal and idler arrival windows, and each data 
point was averaged for 60\,seconds. The normalized histograms are plotted in 
Fig.\,\ref{pump_scan_a} without any background subtraction. For the optimal pump
power ratio ($\sim$360\,mW for downconversion, $\sim$580\,mW for upconversion) 
the maximum singles (coincidence) rate at the center of the
distribution was $\sim$5300/s ($\sim$17/s), including the background. The 
background level in singles counts were $\sim$1900/s for the optimal pump 
power-ratio, corresponding to a background probability per pulse of 
$\sim$2.4$\times 10^{-5}$. The temporal width for singles distribution was 
$\sim$1.3\,ps, consistent with the two-photon coherence time of $\sim$1.4\,ps 
\cite{kuzucu}. Due to the time anti-correlated generation of signal and idler, 
the coincidence profile exhibited a $\sim$165\,fs FWHM width, which is 
significantly narrower than the singles histograms. As the upconversion pulse 
was scanned through the arrival windows of both photons, the only instance where 
the two upconverters could simultaneously detect photons was around the time 
origin.  For an upconversion pump power of 580\,mW, the internal conversion 
efficiency was estimated to be 25\% \cite{ol_submission}.  However, the 
upconversion probability per pump pulse was actually lower because the pump 
pulse was much shorter than the effective pulse width of the signal and idler. 

\begin{figure}[h]
\includegraphics[width=3in]{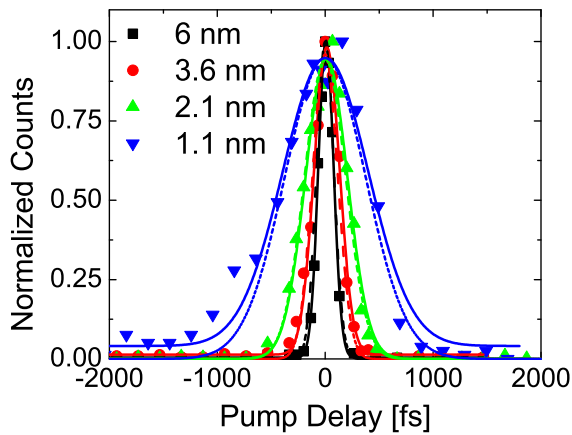} 
\caption{(Color online) Normalized coincidence histograms for various SPDC pump 
3-dB bandwidths: (6-, 3.6-, 2.2-, and 1.1-nm).  Theoretical coincidence profiles 
are plotted as dashed lines.} \label{all_pump_sweeps}
\end{figure}

In order to manipulate the joint temporal amplitude without affecting the 
upconversion timing performance, we modified only the SPDC pump bandwidth by 
inserting a filter from a set of interference filters (3-dB bandwidths: 3.6\,nm, 
2.1\,nm, and 1.1\,nm) before the PPKTP crystal. The measured normalized 
coincidence histograms for different SPDC pump bandwidths are plotted in 
Fig.~\ref{all_pump_sweeps}. As the SPDC pump bandwidth was reduced, the 
single-photon coherence time increased and consequently the coincidence peaks 
became wider. In the same figure, we also show the theoretical predictions for 
the coincidence histograms that we calculate based on the joint temporal 
amplitude with a finite-duration upconversion pump pulse. The parameters for the 
calculation are the upconversion and downconversion pump bandwidths and the 
two-photon coherence bandwidth that we measured with the HOM interference 
\cite{kuzucu}. We assume a flat spectral phase profile in our joint temporal 
amplitude calculation leading to predicted temporal coincidence profiles that 
suggest transform-limited two-photon states.  The good agreement in 
Fig.~\ref{all_pump_sweeps} between data and theory indicates that the SPDC 
output photon pairs were indeed close to the transform limit.  This observation 
is only possible with time-domain measurements because frequency-domain methods 
would be insensitive to dispersive broadening of the photons.

\begin{figure}[h]
\begin{center}
\begin{minipage}{4cm}
    \begin{center}
        \includegraphics[width=4cm]{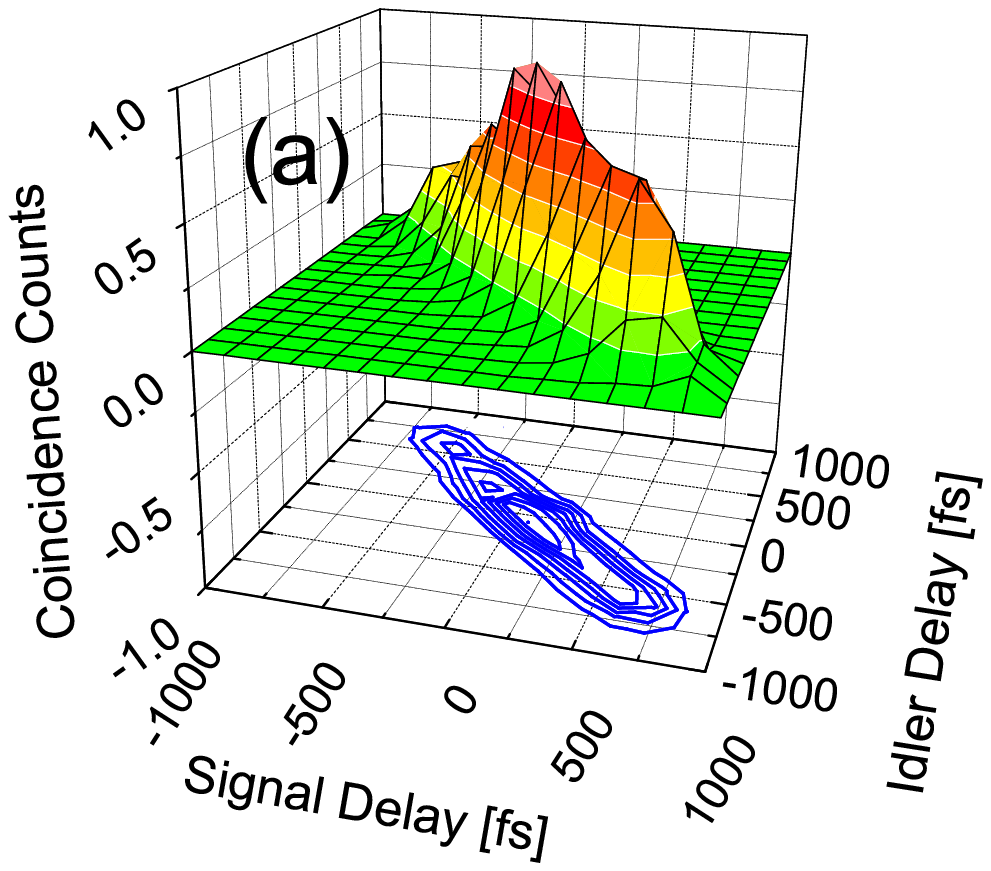}
  \end{center}
\end{minipage}
\
\begin{minipage}{4cm}
    \begin{center}
     \includegraphics[width=4cm]{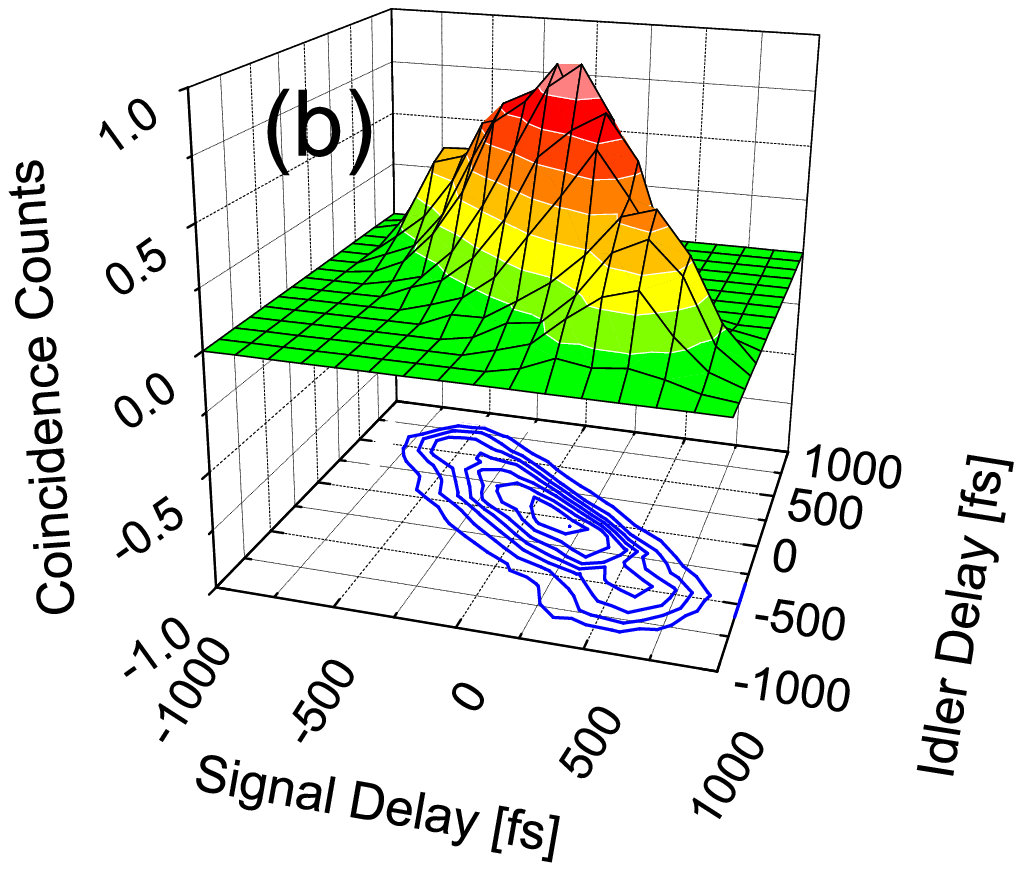}
  \end{center}
\end{minipage}
\
\begin{minipage}{4cm}
    \begin{center}
     \includegraphics[width=4cm]{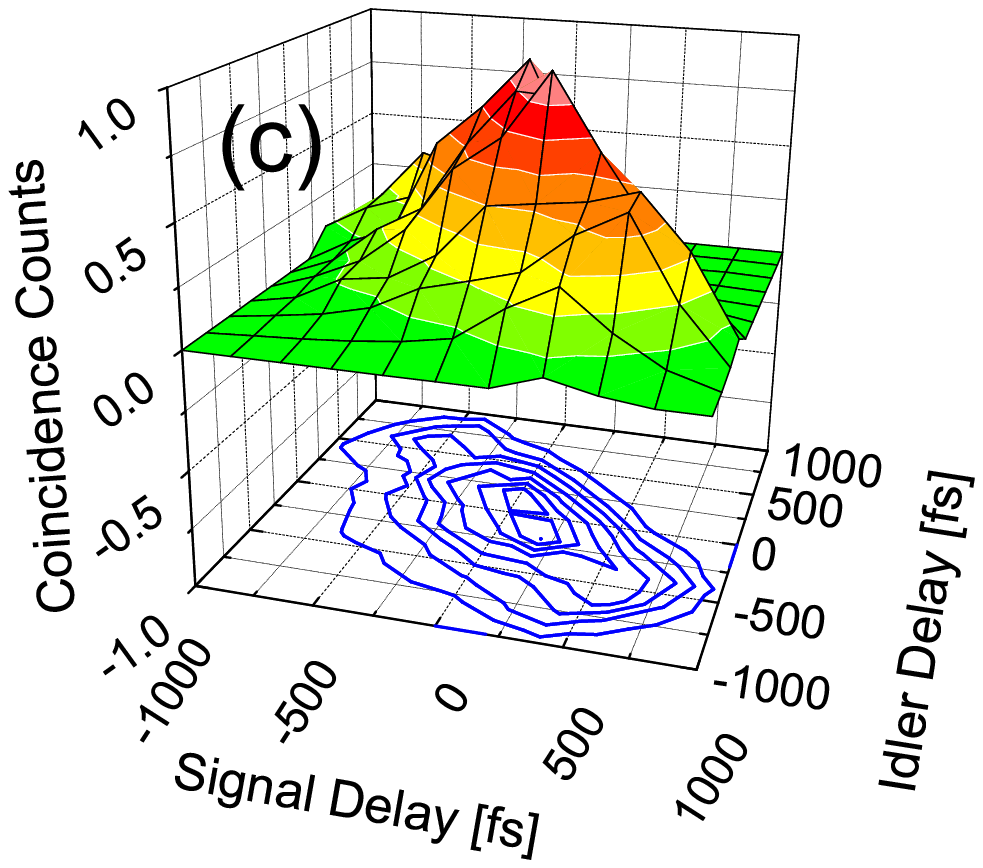}
  \end{center}
\end{minipage}
\
\begin{minipage}{4cm}
    \begin{center}
     \includegraphics[width=4cm]{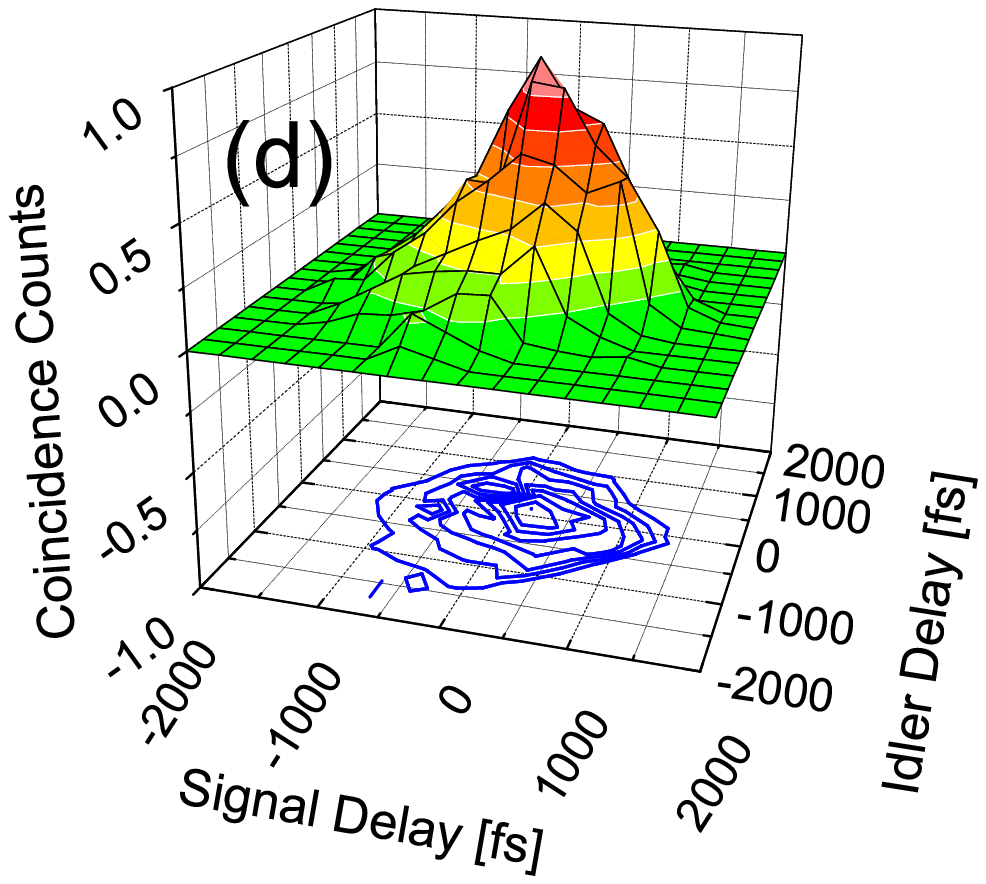}
  \end{center}
\end{minipage}
\end{center}
\caption{(Color online) Experimental joint temporal densities for various 
downconversion pump 3-dB bandwidths: (a)\,6\,nm, (b)\,3.6\,nm, (c)\,2.2\,nm,
(d)\,1.1\,nm. } \label{all_jtds}
\end{figure}

The time-resolved upconversion method enabled us to measure the joint temporal 
density by varying the signal and idler relative delays independently. We set 
the upconversion pump bandwidth to $\sim$6\,nm, and we made the JTD measurements 
using one of the four SPDC pump bandwidths. The coincidence counts were recorded 
over a two-dimensional time grid with 60-s averaging for each data point. For 
all SPDC pump bandwidths except 1.1\,nm, the grid size for the time delays was 
set to 2\,ps$\times$2\,ps (with a 133\,fs step size). We increased the grid size 
to 4\,ps $\times$4\,ps (266\,fs step size) for the 1.1\,nm pump bandwidth. The 
normalized coincidence data for all SPDC pump bandwidths are shown as
surface plots over the two-dimensional time grids in Fig.~\ref{all_jtds}(a)-(d). 
We see dramatic changes in the JTD profile with a change of the SPDC pump 
bandwidth.  With a 6\,nm SPDC pump bandwidth the JTD coincidence profile clearly 
exhibits time anti-correlation that is indicative of two-photon 
coincident-frequency entanglement \cite{kuzucu}. With smaller pump bandwidths, 
the JTD distributions become more symmetric, which corresponds to reduced 
temporal and spectral correlations.


\begin{figure}[htb]
\centerline{\includegraphics[width=3in]{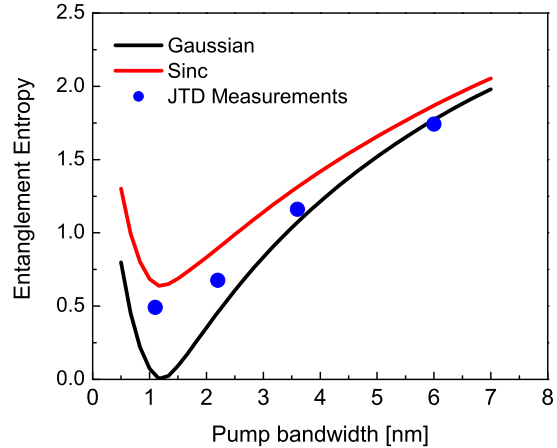}} \caption{(Color 
online) Entanglement entropy values calculated from experimental JTD 
distributions for various SPDC pump bandwidths of Fig.~4. The theoretical 
entropy variations for Gaussian (black) and sinc-type (red) phase-matching
functions are given in solid curves.}\label{entropy}
\end{figure}

We can quantify the two-photon frequency entanglement as a function of the pump 
bandwidth based on the measured JTD distributions and by using Schmidt 
decomposition for continuous variables \cite{law}.  In this formalism, the joint 
temporal amplitude, ${\cal{A}}(\tau_s,\tau_i)$, is expressed as a discrete sum 
of the temporal eigenmodes with eigenvalues $\lambda_n$, through which the
entanglement entropy can be computed as $S=-\sum_{k=1}^n \lambda_k \log_2 
\lambda_k$ \cite{law}. Figure~\ref{entropy} shows the computed entanglement 
entropy from the experimental JTD distributions in Fig.~4 assuming that the 
joint state is transform limited. For comparison, we have also calculated the 
theoretical entropy curves as a function of the SPDC pump bandwidth, where the 
pump spectrum is assumed to be Gaussian. Two curves are plotted in 
Fig.~\ref{entropy}, one representing a Gaussian and the other a sinc 
phase-matching function.  For a Gaussian phase-matching function, a fully  
factorizable two-photon state is predicted with a pump bandwidth
of $\sim$1.2\,nm, and yielding an entropy of zero. For the more realistic sinc 
function for the phase matching, a highly but not completely factorizable 
two-photon state is achievable. Since the sinc-type spectral response 
corresponds to a boxcar shape in the time domain, it necessitates the inclusion 
of higher order Schmidt modes and hence increases the entanglement entropy.

Figure~\ref{entropy} shows a good qualitative agreement between the theoretical 
entropy curves and the entropy values obtained from the JTD distributions. The 
entanglement entropy corresponding to the experimental JTD profiles are lower 
than the theoretical curve for the sinc phase-matching function.  This is 
reasonable if we take into account that the actual time-domain profile of the 
phase-matching function is smoother than a boxcar shape because of grating 
inhomogeneity, as confirmed by the singles histogram measurements of 
Fig.\,\ref{pump_scan_a}. Therefore, the experimental JTD distributions can be 
expressed with a smaller number of Schmidt modes, resulting in a lower 
entanglement entropy than that of the theoretical of a sinc function. For a 
1.1-nm SPDC pump bandwidth, which yields an output that is nearly factorizable, 
we have computed the purity of the heralded single-photon state as $\sim$0.88, 
where purity is defined as $p={\rm Tr}({\hat{\rho}}^2_S)=\sum_{n=0}^{\infty} 
\lambda^2_n$ \cite{law, uren}. This purity value compares well with that of the 
pure-state single photons generated under SPDC using a different spectral 
engineering method \cite{mosley}.  We believe that the purity can be further 
improved by finer control over the pump bandwidth and additional spectral
filtering. In comparison, the output for the case of a 6-nm SPDC pump bandwidth 
yields a purity of $\sim$0.38, which is a consequence of the high degree of 
coincident-frequency entanglement. 

In conclusion, we have developed a time-domain measurement technique for single 
photons with sub-picosecond resolution that we used to measure the two-photon 
joint temporal density for the first time.  We applied the technique to verify  
anti-correlation in the arrival times of the signal and idler photons that were
coincident-frequency entangled.  Finally, the new tool allowed us to monitor the 
effect of varying the SPDC pump bandwidths, leading to the generation of a 
nearly factorizable two-photon state, which should be of interest to many 
quantum information processing applications. We believe that the JTD measurement 
technique is a powerful tool for engineering temporal and spectral correlations 
of ultrafast SPDC photons. Such a characterization technique would complement 
the frequency-domain counterparts to quantify and manipulate multi-photon 
entanglement for quantum information processing applications.

\begin{acknowledgments}
This work was supported in part by the Hewlett-Packard Laboratories and by the 
National Institute of Information and Communications Technology, Japan.
\end{acknowledgments}



\begin{thebibliography}{99}
\bibitem{klm01}E. Knill, R. Laflamme, and G. J. Milburn, Nature (London) {\bf 
409,} 46 (2001).
\bibitem{grice} W. P. Grice and I. A. Walmsley, Phys. Rev. A {\bf 56,} 1627 
(1997); W. P. Grice, A. B. U'Ren, and I. A. Walmsley {\em ibid.} {\bf 64,} 
063815 (2001).
\bibitem{giovannetti_epm}V. Giovannetti, L. Maccone, J. H. Shapiro, and F. N. C. 
Wong, Phys. Rev. Lett. {\bf 88,} 183602 (2002); Phys. Rev. A {\bf 66,} 043813
(2002).
\bibitem{kuzucu} O. Kuzucu \emph{et al.}, Phys. Rev. Lett. {\bf 94}, 083601 
(2005).
\bibitem{giovannetti_nature}V. Giovannetti, S. Lloyd, and L. Maccone, Nature 
(London) {\bf 413,} 417 (2001);  Science {\bf 306,} 1330 (2004).
\bibitem{uren} A. B.\ U'Ren \emph{et al.}, Laser Phys. {\bf 15}, 146 (2005).
\bibitem{mosley} P. J. Mosley \emph{et al.}, Phys. Rev. Lett. {\bf 100}, 133601 
(2008).
\bibitem{hendrych} M. Hendrych, M. Micuda, and J. P. Torres, Opt. Lett. {\bf 
32,} 2339 (2007); A. Valencia \emph{et al.}, Phys. Rev. Lett. {\bf 99,} 243601
(2007).
\bibitem{HOM}C. K. Hong, Z. Y. Ou, and L. Mandel, Phys. Rev. Lett. {\bf 59,} 
2044 (1987).
\bibitem{wasilewski}W. Wasilewski, P. Kolenderski, and R. Frankowski, Phys. Rev. 
Lett. {\bf 99,} 123601 (2007).
\bibitem{ol_submission} O. Kuzucu, F. N. C. Wong, S. Kurimura, and S. Tovstonog, 
submitted to \emph{ Opt. Lett.}
\bibitem{law} C. K. Law, I. A. Walmsley, and J. H. Eberly, Phys. Rev. Lett. {\bf 
84,} 5304 (2000); S. Parker, S. Bose, M. B. Plenio, Phys. Rev. A {\bf 61,} 
032305 (2000); L. Lamata and J. Le\'{o}n, J. Opt. B {\bf 7,} 224 (2005).

\end{thebibliography}
\end{document}